\documentclass[a4paper,fleqn,usenatbib]{mnras}

\usepackage{newtxtext,newtxmath}

\usepackage[T1]{fontenc}
\usepackage{ae,aecompl}

\usepackage{graphicx}	
\usepackage{amsmath}	
\usepackage{amssymb}	
\usepackage{color}
\usepackage[english]{babel}
\usepackage{graphicx}
\usepackage{hyperref}
\usepackage{listings}
\usepackage{lscape}
\usepackage{natbib}
\usepackage{url}
\usepackage{breakurl}
\usepackage{xspace}
\bibpunct{(}{)}{;}{a}{}{,}

\newcounter{Rco}
\newcommand{\ionw}[3]{\mbox{\ion{#1}{#2}~$\lambda\,#3\,\mathrm{\AA}$}\xspace}

\newcommand{\Jonw}[3]{\mbox{\ion{#1}{#2}~$\lambda\,#3$\,\AA}\xspace}
\newcommand{\Jonww}[3]{\mbox{\ion{#1}{#2}~$\lambda\lambda\,#3$\,\AA}\xspace}

\newcommand{\logg}{\mbox{$\log g$}\xspace}
\newcommand{\loggw}[1]{\mbox{$\log g\hspace{-0.5mm} =\hspace{-0.5mm}  #1$}}

\newcommand{\Teff}{\mbox{$T_\mathrm{eff}$}\xspace}
\newcommand{\Teffw}[1]{\mbox{$\Teff\hspace{-0.5mm} =\hspace{-0.5mm} #1 \,\mathrm{K}$}}

\newcommand{\Msol}{$M_\odot$}

\newcommand{\mmspr}{\hbox{}\hspace{+0.8cm}}
\newcommand{\smspr}{\hbox{}\hspace{+2.5mm}}

\newcommand{\pn}{PN\,PRTM\,1\xspace}
\title[Stellar parameters for the central star of the planetary nebula PRTM\,1]
      {Stellar parameters for the central star of the planetary nebula PRTM\,1 using the German Astrophysical Virtual Observatory service TheoSSA}

\author[T\@. Rauch et al.]{
T\@. Rauch$^{1}$\thanks{E-mail: gavo@listserv.uni-tuebingen.de},
M\@. Demleitner$^{2}$\thanks{E-mail: gavo@ari.uni-heidelberg.de},
D\@. Hoyer$^{1}$,
and 
K\@. Werner$^{1}$
\\
$^{1}$Institute for Astronomy and Astrophysics,
           Kepler Center for Astro and Particle Physics,
           Eberhard Karls University, 
           Sand 1,
           72076 T\"ubingen, 
           Germany\\
$^{2}$Astronomisches Rechen-Institut (ARI), 
           Centre for Astronomy of Heidelberg University, 
           M\"onchhofstra\ss e 12-14, 
           69120 Heidelberg, 
           Germany
}

\date{Accepted 2018 January 2. Received 2017 December 12; in original form 2017 November 14}

\pubyear{2017}

\begin{document}
\label{firstpage}
\pagerange{\pageref{firstpage}--\pageref{lastpage}}
\maketitle

\begin{abstract}
The German Astrophysical Virtual Observatory (GAVO) developed the registered service 
TheoSSA (theoretical stellar spectra access) 
and the supporting registered VO tool 
TMAW (T\"ubingen Model-Atmosphere WWW interface). 
These allow individual spectral analyses of hot, compact stars with state-of-the-art 
non-local thermodynamical equilibrium (NLTE) stellar-atmosphere models that presently
consider opacities of the elements H, He, C, N, O, Ne, Na, and Mg, without requiring 
detailed knowledge about the involved background codes and procedures.
Presently, TheoSSA provides easy access to about 150\,000 pre-calculated stellar SEDs and is
intended to ingest SEDs calculated by any model-atmosphere code. 
In the case of the exciting star of \pn, we demonstrate the easy way to calculate individual NLTE
stellar model-atmospheres to reproduce an observed optical spectrum.
We measured
\Teffw{98\,000\pm 5\,000},
$\log\,(g\,/\,\mathrm{cm/s^2}) = 5.0^{+0.3}_{-0.2}$, 
and photospheric mass fractions of
H  $= 7.5 \times 10^{-1}$ (1.02 times solar),
He $= 2.4 \times 10^{-1}$ (0.96),
C  $= 2.0 \times 10^{-3}$ (0.84),
N  $= 3.2 \times 10^{-4}$ (0.46),
O  $= 8.5 \times 10^{-3}$ (1.48) with uncertainties of $\pm 0.2$\,dex.
We determined the stellar mass and luminosity of 
$0.73^{+0.16}_{-0.15}\,M_\odot$ and
$\log (L /  L_\odot) = 4.2 \pm 0.4$, respectively. 
\end{abstract}




\section{Introduction}
\label{sect:intro}

\begin{figure}
  \resizebox{\hsize}{!}{\includegraphics{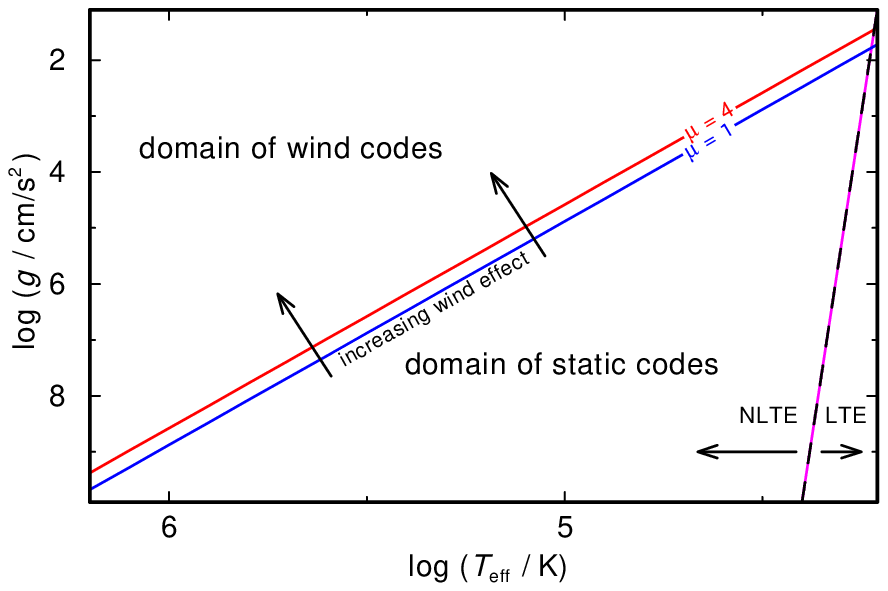}}
  \caption{Approximate locations of the domains of LTE and NLTE stellar-atmosphere models 
           for static and expanding stellar atmospheres.
           The diagonal lines indicate Eddington limits for different mean atomic weights
           $\mu$ of the atmosphere.} 
  \label{fig:TheoSSA_edd}
\end{figure}

For precise spectral analysis of hot stars, observations with good quality and
stellar-atmosphere models that account for reliable physics and 
deviations from the assumption of a local thermodynamic equilibrium (LTE)
are necessary. Figure\,\ref{fig:TheoSSA_edd} illustrates that for cool 
stars with high surface gravity (spectral type B and later),
LTE model atmospheres might be sufficient \citep[cf\@.,][]{auermihalas1972}. 
There are, however, non-local thermodynamic equilibrium
(NLTE) effects in any star, at least towards higher energies and in 
high-resolution spectra, particularly in the infrared (IR) wavelength range. 

In the last century, NLTE model atmospheres were believed to be a domain of specialists, because
they require high computational times and a lot of work on atomic data. Hence, the use of 
blackbody spectra to represent hot stars had been a common approximation in work
where ionizing fluxes were required.

Several well documented NLTE codes were developed in that time \citep{hubenymihalas2014}, e.g.,
the PHOENIX  \citep[\url{http://www.hs.uni-hamburg.de/EN/For/ThA/phoenix}][]{baronetal2010,hauschildtbaron1999,hauschildtbaron2010},
the TLUSTY   \citep[\url{http://nova.astro.umd.edu}][]{hubenylanz2011,hubeny1988,hubenylanz1992,hubenyetal1994,hubenylanz1995},
and the TMAP \citep[http://www.uni-tuebingen.de/de/41621,][]{tmap2012} code.
The latter two are widely used, e.g., in spectral analyses of hot, compact stars. TMAP was successfully used for
sdO/B stars
\citep[e.g.,][]{
rauchetal1991,rauch1993,rauchetal2002,rauchetal2010b,klepprauch2011,rauchetal2014},
PG\,1159-type stars
\citep[e.g.,][]{
werneretal1991,jahnetal2007,werneretal2015,werneretal2016},
DA-type white dwarfs
\citep[e.g.,][]{
wernerrauch1997,rauchetal2013},
DO-type white dwarfs
\citep[e.g.,][]{
hoyeretal2017,rauchetal2017,wernerrauch1997},
central stars of planetary nebulae
\citep[CSPN, e.g.,][]{
ercolanoetal2003,rauchetal2007,ziegleretal2012}, 
super-soft X-ray sources
\citep[e.g.,][]{
vanteeselingetal1999,rauchetal2010b,orioetal2013,peretzetal2016}, and
neutron stars
\citep[e.g.,][]{
rauchetal2008}.

In the framework of the Virtual Observatory (VO), the 
German Astrophysical Virtual Observatory (GAVO\footnote{\url{http://www.g-vo.org}}) project nowadays provides 
synthetic stellar spectra on demand via the registered
Theoretical Stellar Spectra Access VO service 
\citep{rauch2008a,rauchnickelt2009,rauchetal2009}. TheoSSA is designed to host
spectral energy distributions (SEDs) calculated by any model-atmosphere code.
Initially, these SEDs should be used as realistic stellar ionizing fluxes for
photoionization models of planetary nebulae (PNe).
In addition, such SEDs can be used for spectral analyses
\citep{rauchetal2010a,ringatrauch2010,rauchringat2011,ringatrauchwerner2012,ringatPhD2013}. The registered 
VO tool TMAW\footnote{\url{http://astro.uni-tuebingen.de/~TMAW}} 
(T\"ubingen NLTE Model-Atmosphere WWW Interface)
allows to calculate model atmospheres for hot, compact stars with the
T\"ubingen NLTE model-atmosphere package
\citep[TMAP\footnote{\url{http://astro.uni-tuebingen.de/~TMAP}}]{rauchdeetjen2003,werneretal2003},
considering opacities of H, He, C, N, O, Ne, Na, and Mg (with individually user-chosen abundances).
To represent these species, TMAW uses either standard or individually compiled model atoms based on data provided
by the T\"ubingen Model Atom Database (TMAD\footnote{\url{http://astro.uni-tuebingen.de/~TMAD}},
cf., TMAP User's Guide \url{http://astro.uni-tuebingen.de/~TMAP/UserGuide/UserGuide.pdf}).
TMAW SEDs allow individual, detailed analyses and a more
precise determination of effective temperature (\Teff), surface gravity (\logg), and photospheric
abundances of a star. Once calculated, the TMAW SEDs are automatically ingested by TheoSSA.
Tables\,\ref{lst:request} and \ref{lst:results} show a TMAW request and the files that can be
retrieved by the TMAW user after the end of the calculation, respectively.

The primary purpose of these SEDs was to perform preliminary spectral analyses and to achieve determinations
of \Teff, \logg, and abundances better than 20\%. \citet{rauchringat2012} and \citet{ringatreindl2013} 
demonstrated that the deviation of \Teff and \logg values found with TMAW SEDs and with 
``manually calculated'' TMAP models is of the order of 10\%. To demonstrate a spectral analysis using
TheoSSA and TMAW, we have chosen the hot central star (CS) of the planetary nebula PRTM\,1
because no reliable determination of photospheric parameters like \Teff, \logg, and element abundances
is hitherto available.

\citet{torrespeimbertetal1990} discovered the very-high excitation PN PRTM\,1
(PN\,G243.8$-$37.1), located in the Galactic halo at a distance of $d = 5 \pm 2\,\mathrm{kpc}$
and at a height below the Galactic plane of $|z| = 3 \pm 1.2\,\mathrm{kpc}$. 
\citet{penaetal1990} found that, although carbon-poor, its heavy elements are
not as underabundant as known from other halo PNe.
In their PN analysis, they used a grid of SEDs from NLTE H+He models \citep{cleggmiddlemass1987} and 
best reproduced the nebular ionization structure using an ionizing source with 
\Teff $\approx 90\,000\,\mathrm{K}$ and a surface gravity of $\log\,(g\,/\,\mathrm{cm/s^2}) \approx 5.3$.
This was in agreement with \Teffw{90\,000 \pm 10\,000} determined by \citet[from blackbody fits to the stellar continuum flux]{feibelman1998}
and preliminary results of M\'endez \& Ruiz \citep[priv\@. comm\@., mentioned by][analysis method and
error ranges unknown, based on optical observations]{penaetal1990} 
of $T_\mathrm{eff} \approx 80\,000\,\mathrm{K}$ and $\log g \approx 5.2$ that were then used to determine the stellar mass of $M = 0.58\,M_\odot$.
\citet{penaetal1990} discovered a strong C deficiency of the order of 1\,dex in the PN
and suggested that the progenitor star was metal-poor with very low mass that did not undergo 
a third dredge up.

\citet{mendez1991} classified the CS of \pn as an O(H)-type star, i.e., 
its optical spectrum is dominated by hydrogen absorption lines. 
Stellar spectroscopic and photospheric variability was reported by
\citet[in the ultraviolet]{feibelman1998} and was confirmed by
\citet[in the optical with $\Delta m_\mathrm{V} \la 1\,\mathrm{mag}$,
$m_\mathrm{V} = 16.6$ in Oct 1990 and
$m_\mathrm{V} = 15.6 \pm 0.2$ in Feb 1998]{penaruiz1998}.

Recently, the CS had been monitored during three months for its variability by \citet{boffinetal2012a},
because its spectroscopic appearance matches that of the CSPN of Fleming\,1
\citep[PN\,Fg\,1, PN\,G290.5+07.9,][]{boffinetal2012a} and thus,
a binary nature was hypothesized. 
The obtained data, however, does not support this.
The CS's rectified average 
FORS2\footnote{FOcal Reducer and low-dispersion Spectrograph} spectrum 
(4597\,\AA\ -- 5901\,\AA, 
slit width 0\farcs 5, 
volume-phase holographic grating 1400V,
resolution 1.22\,\AA\ full width at half maximum) 
taken at ESO's VLT\footnote{European Southern Observatory, Very Large Telescope}
is of good quality, suitable for a spectral analysis.
It exhibits a variety of lines, namely of
\ion{H}{i}, \ion{He}{ii}, \ion{C}{iv}, \ion{N}{v}, \ion{O}{v}, and \ion{O}{vi}. 
The latter two even allow to evaluate the O ionization equilibrium to determine \Teff precisely.

Since hitherto no spectral analysis appeared in the literature, 
we take the opportunity and use the average spectrum of this spectroscopically variable star
to demonstrate how straightforward a NLTE spectral analysis became with the development of TheoSSA and TMAW.
A time-dependent spectral analysis is desirable to further investigate on the nature
of the star. This, however, is not the focus of this paper.
 
We start with a description of TheoSSA in Sect.\,\ref{sect:theossa}.
In Sect.\,\ref{sect:prelim}, 
we briefly describe the coarse determination of \Teff and \logg using pre-calculated synthetic H+He SEDs. 
A more individual analysis based on newly calculated stellar atmosphere models and SEDs is presented in Sect.\,\ref{sect:tmaw}. 
We summarize our results and conclude in Sect.\,\ref{sect:results}.

\section{TheoSSA}
\label{sect:theossa}

TheoSSA is a spectral service fully integrated into the VO. This means that

\begin{enumerate}
\item TheoSSA can easily be located using the VO Registry \citep{paper:regclient}, a
global inventory of services allowing the discovery of data services
using a wide variety of constraints.  Three example queries that would,
among others, yield TheoSSA, are ``all spectral services serving model
spectra in the ultraviolet'', ``services dealing with white dwarfs'',
and ``services created by Thomas Rauch''.  The metadata provided by the
Registry also contains information on who to contact in case of
technical problems or scientific questions, and what bibliographic sources
to cite when using the data served.  By virtue of being in the Registry,
TheoSSA has a special URI\footnote{unified resource identifier}, the IVOID \citep{std:VOID}, that
uniquely identifies it: \nolinkurl{ivo://org.gavo.dc/theossa/q/ssa}.

\item TheoSSA can be queried using a standard protocol called Simple
Spectral Access \citep[SSA,][]{std:SSAP}.  This means that computer
programs (``clients'') can use the same code to query TheoSSA and
numerous other spectral services which are also discoverable through the
VO Registry.  Examples for such clients include Splat
\citep{soft:Splat} and VOSpec \citep{soft:VOSpec}, both containing
functionality geared towards spectral analysis, as well as 
TOPCAT\footnote{Tool for OPerations on Catalogues And Tables}
\citep{soft:Topcat} for generic processing of tabular data (which
includes spectra).  For use without VO-enabled client software, TheoSSA
also provides a conventional form-based interface for use with web 
browsers\footnote{\url{http://dc.g-vo.org/theossa}}.

\item TheoSSA delivers spectra in standard formats, in particular in
VOTables following the Spectral Data Model \citep{std:SDM}.  This lets
clients not only immediately display spectra obtained from the service
endpoint. The rich metadata also allows automatic unification of
properties like units or spectral resolution between spectra obtained
from different sources.

\item TheoSSA offers an associate prototype service implementing
the upcoming SODA (``Service Operations On Data'') standard.  This
allows on-the-fly, server-side format conversion and cutouts in the
context of the general Datalink protocol, which itself is a generic way
to link datasets with related artifacts \citep{std:Datalink}.

\end{enumerate}

TheoSSA is operated at GAVO's Heidelberg data center on top of the
publicly available VO server package DaCHS \citep{soft:DaCHS}.

SSA, the VO's standard
protocol for querying spectral services like TheoSSA, is in its core
strongly biased towards observational data.  Typical
parameters used there include location of the aperture projected at the sky or
the name of the target object.  Though a few spectra in TheoSSA were
computed assuming abundances of well-known objects (see below) and can be located by
searching for these objects' names, typically such parameters are
not useful to constrain results from a ``theory'' service
(in VO jargon, this denotes services publishing computed, rather than
observed, data).

Instead, a service like TheoSSA employs numerous custom parameters, in
particular \texttt{w\_<element>} where \texttt{<element>} is an element
name for the model abundances, \texttt{t\_eff} for the model effective
temperature, \texttt{mdot} for the mass loss rate, and \texttt{log\_g}
for the logarithm of the surface gravity (Figs.\,\ref{fig:TheoSSA_input} and
\ref{fig:TheoSSA_output} display the TheoSSA TMAP Web Interface and its result page,
respectively).  All these float-valued parameters
allow the specification of ranges using a slash; for instance
\texttt{log\_g=5.5/7} will select all spectra with $5.5\leq \log g \leq
7$, with half-open variants like \texttt{/7} or \texttt{5.5/} working as
expected.  This is in keeping with the syntax of the core SSA
parameters. The SSA standard allows declaring such custom parameters,
and TheoSSA does so to enable its operation from clients such as
Splat.

The well-known objects mentioned above in particular include stellar flux 
standards like
the DA-type white dwarf (DA) EG\,274,
the OB-type subdwarf (sdOB) Feige\,24 \citep{rauchetal2014},
Feige\,67 (sdO),
Feige\,110 (sdO),
GD\,50 (DA),
GD\,71 (DA),
GD\,108 (sdB),
GD\,153 (DA),
G191$-$B2B \citep[DA,][]{rauchetal2013},
G\,93$-$48 (DA),
HZ\,2 (DA),
HZ\,43 (DA), and
Sirius\,B (DA).
Spectra of these standard stars were computed for the full range of 
3000 through 55\,000\,\AA.

TheoSSA is not limited to publishing spectra from TMAP. Researchers
producing theoretical spectra are cordially invited to contact the
authors for further information on how to have their spectra published
in TheoSSA, too.

\section{Preliminary spectral analysis: TheoSSA}
\label{sect:prelim}

The cookbook recipe for spectral analysis reads easy. Firstly find a reasonable guess for
\Teff and \logg. Secondly include all (at least) identified species
and adjust their abundances to reproduce their observed line strengths. Thirdly fine-tune
\Teff, \logg, and abundances to improve the agreement between observation and model.
In the following, we will apply this recipe to observations of the CS of 
\pn.

\begin{figure}
  \resizebox{\hsize}{!}{\includegraphics{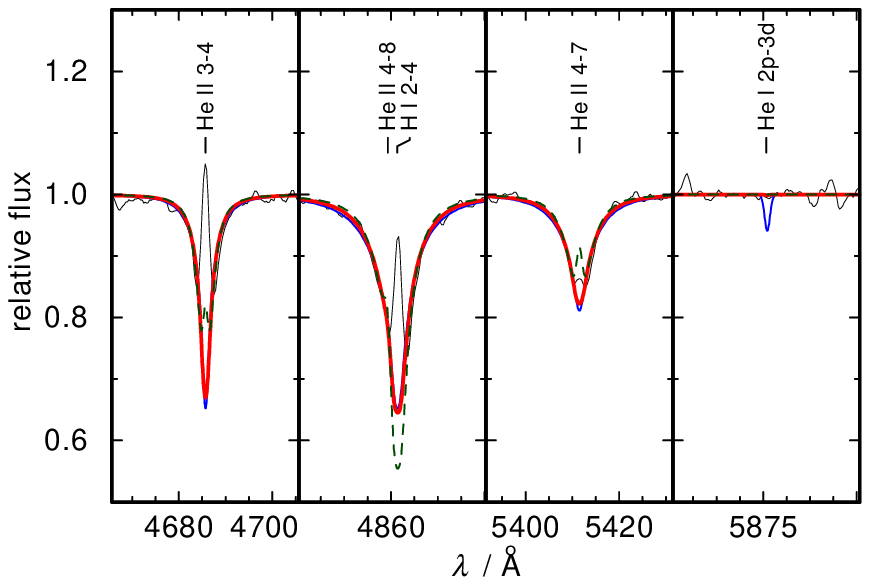}}
  \caption{Four sections of the FORS2 observation of the central star of \pn
           (all observations shown in this
           paper are processed with a \citet{savitzkygolay1964} low-band filter)
           compared with synthetic spectra (convolved with a Gaussian to simulate
           the FORS2 resolution, see Sect.\,\ref{sect:intro}) calculated 
           from H+He-composed models with
          \Teff = 60\,000\,K (blue, thin), 80\,000\,K (red, thick), and 100\,000\,K (green, dashed),
           $\log g = 5.0$, and solar abundances \citep{asplundetal2009}.
           The central emissions in the \ion{H}{i} and \ion{He}{ii} line cores are of nebular origin.
          } 
  \label{fig:TheoSSA_coarse_Teff}
\end{figure}

\begin{figure}
  \resizebox{\hsize}{!}{\includegraphics{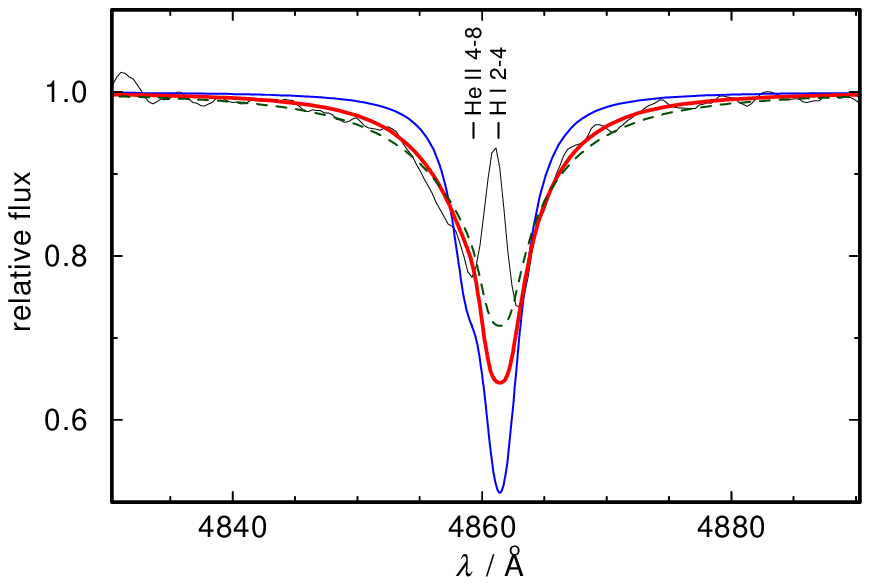}}
  \caption{Section of the FORS2 observation around H\,$\beta$  compared with synthetic spectra 
           calculated from H+He-composed atmosphere models with
           \loggw{4.5} (blue, thin), 5.0 (red, thick), and 5.5 (green, dashed),
           \Teffw{80\,000}, and solar abundances.
          } 
  \label{fig:TheoSSA_coarse_logg}
\end{figure}

To find a start model for our further calculations that include metal opacities, we retrieved
SEDs of H+He composed models from TheoSSA. 
In Figure 2, the
FORS2 observations of the central star are shown. The large intensity
of \ionw{He}{ii}{4685.69} and the absence of 
\ionw{He}{i}{5875.62} (2p\,$^3$P$^\mathrm{o}$ -- 3d\,$^3$D) in the FORS2 observation indicates
\Teff $\ga 60\,000$\,K. Figure\,\ref{fig:TheoSSA_coarse_Teff} shows that this line fades in the model
at about \Teffw{70\,000}. The line core of \ionw{He}{ii}{5411.53} becomes too shallow at
\Teff $\ga 90\,000$\,K. This yields a preliminary \Teffw{80\,000 \pm 15\,000}. 
The selected \loggw{5.0} reproduces the line wings of the \ion{H}{i} and \ion{He}{ii} 
lines\footnote{Stark line-broadening tables of 
\citet[][extended tables of 2015, priv\@. comm\@.]{tremblaybergeron2009} and 
\citet{schoeningbutler1989}
are used to calculate the theoretical \ion{H}{i} and \ion{He}{ii} line profiles, respectively.
}.
Figure\,\ref{fig:TheoSSA_coarse_logg} verifies this in detail for H\,$\beta$. 
Its theoretical line profile has significantly too-narrow outer wings at \loggw{4.5}, while
they are slightly broader than observed at \loggw{5.5}. Although the line core is contaminated 
by residual nebular emission (Fig.\,\ref{fig:TheoSSA_coarse_Teff} shows other nebular lines in
the observation), it is obvious that the inner line core cannot be matched at \loggw{5.5}.
The H / He abundance ratio is solar because at a higher ratio,
the \ionw{He}{ii}{4859.31}\,/\, H\,$\beta$ blend is too broad and at a lower ratio, it is
too narrow.

We adopt \Teffw{80\,000 \pm 15\,000}, \loggw{5.0^{+0.3}_{-0.2}}, and a mass ratio H / He $= 2.96 \pm 0.20$ 
for our further analysis. The values will be improved in Sect.\,\ref{sect:tmaw}.
To achieve these results, only a small expenditure of time of the order of one hour was necessary.

\section{Detailed spectral analysis: TMAW}
\label{sect:tmaw}

\begin{figure*}
  \resizebox{\hsize}{!}{\includegraphics{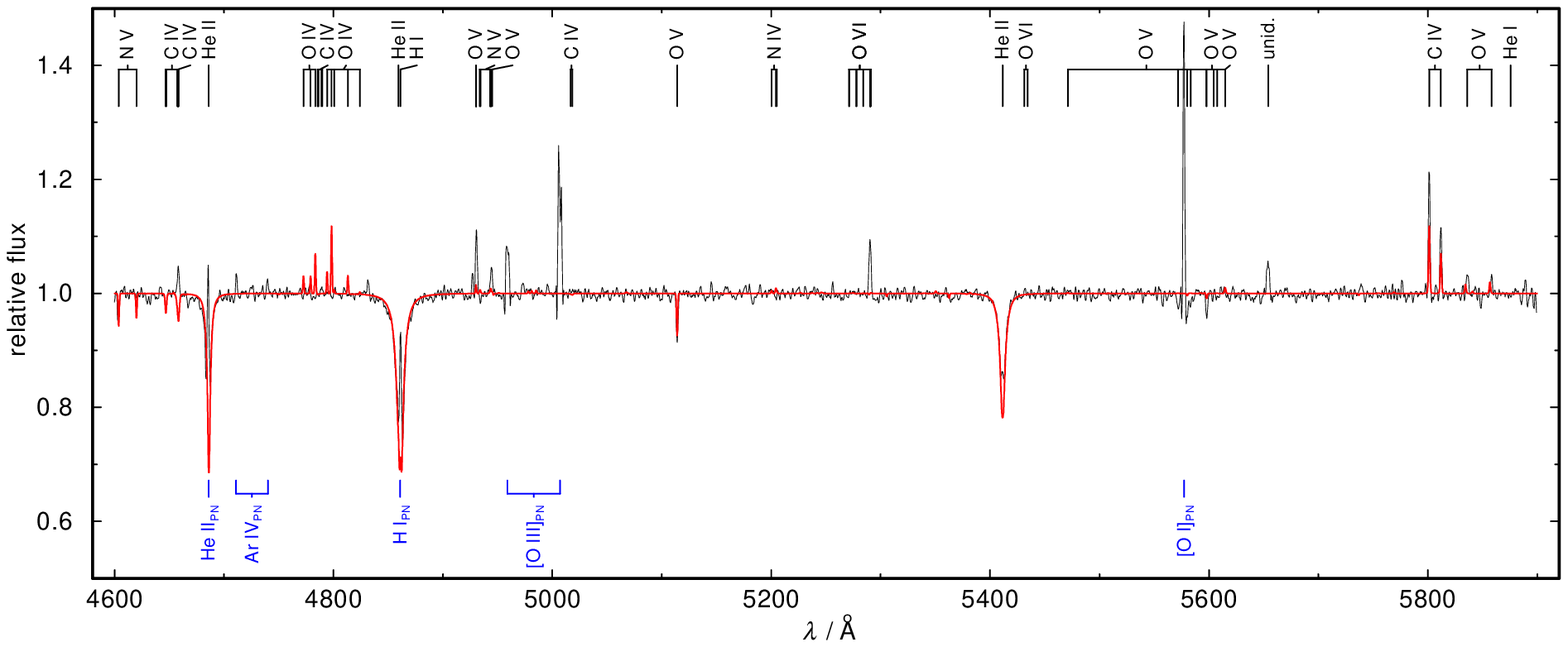}}
  \caption{FORS2 observation 
           compared with a synthetic spectrum calculated 
           from a H+He+C+N+O model with
           \Teffw{80\,000},
           \loggw{5.0}, and solar abundances.
           Identified stellar lines are marked at top, nebular lines at the bottom.
           ``unid.'' denotes an unidentified line.
          } 
  \label{fig:TheoSSA_solar}
\end{figure*}

For a first model (\Teffw{80\,000}, \loggw{5.0}), 
we submitted a TMAW request (Figs.\,\ref{fig:TMAW_input}, \ref{fig:TMAW_output}) which includes 
H, He, C, N, and O with solar abundances.
The comparison of the preliminary synthetic spectrum with the observation (Fig.\,\ref{fig:TheoSSA_solar})
shows results for C, N, and O as follows.\\
{\large\sc Carbon.} 
\Jonww{C}{iv}{5801.33, 5811.98} (3s\,$^2$S$_{1/2}$ -- 3p\,$^2$P$^\mathrm{o}_{3/2}$,
                                 3s\,$^2$S$_{1/2}$ -- 3p\,$^2P$$^\mathrm{o}_{1/2}$) 
are in emission like observed, but too weak.
\Jonw{C}{iv}{4646} (5d\,--\,6f) appears in absorption and is well reproduced.
The observed \Jonww{C}{iv}{4657\,-\,4658} (5f\,--\,6g, 5g\,--\,6h) emission is not
matched, because the model shows an absorption line.\\
{\large\sc Nitrogen.}
\Jonww{N}{v}{4603.73, 4619.98} (3s\,$^2$S$_{1/2}$ -- 3p\,$^2$P$^\mathrm{o}_{3/2}$,
                                3s\,$^2$S$_{1/2}$ -- 3p\,$^2$P$^\mathrm{o}_{1/2}$) 
appear in absorption like observed. The first line is too weak compared with the observation.
\Jonww{N}{v}{4943\,-\,4945} (6f\,--\,7g, 6g\,--\,7h, 6h\,--\,7i, 6h\,--\,7g, 6g\,--\,7f)
is too weak compared with the observed absorption feature.
It is used to determine the N abundance (Fig.\,\ref{fig:TheoSSA_best}).\\
{\large\sc Oxygen.}
Several \ion{O}{v} lines are present in the observation, the strongest emission lines
are
\Jonww{O}{v}{4930.21 - 4930.31} (6h\,$^3$H$^\mathrm{o}$ -- 7i$^3$I)
and
\Jonww{O}{v}{5836.33, 5858.58, 5858.61} (4f'\,$^3$G -- 7h$^3$H$^\mathrm{o}$),
the latter are weaker than observed.
\Jonw{O}{v}{5114.06} (3s\,$^1$S$_{0}$ -- 3p$^1$P$^\mathrm{o}_{1}$)
is an absorption line that matches the observation well.
\Jonww{O}{vi}{5289.51\,-\,5292.07} (7g\,--\,8h, 7f\,--\,8g, 7i\,--\,8h, 7h\,--\,8k, 7h\,--\,8g, 7g\,--\,8f)
is a prominent emission line in the observation, but is not visible in the synthetic spectrum.

From the absence of \Jonw{O}{vi}{5290}, which is commonly used as a strategic line for
\Teff determination and the too-strong \ion{O}{v} emission lines, we estimate that
the degree of ionization in the model is too low, i.e., either \Teff is too low
or \logg is too high. The same is indicated by the too-weak \Jonww{C}{iv}{5801, 5811} 
emission doublet. Since \loggw{5.0} well reproduces the line wings of the \ion{H}{i}
Balmer lines, we calculated additional TMAW models with higher \Teff and
$80\,000\,\mathrm{K} \le  T_\mathrm{eff} \le 105\,000\,\mathrm{K}$ with $\Delta T_\mathrm{eff} = 1000\,\mathrm{K}$.
and kept \logg as well as the abundances fixed.
A screenshot of the respective TMAW request WWW interface is shown in Fig.\,\ref{fig:TMAW_input}.

\begin{figure}
  \resizebox{\hsize}{!}{\includegraphics{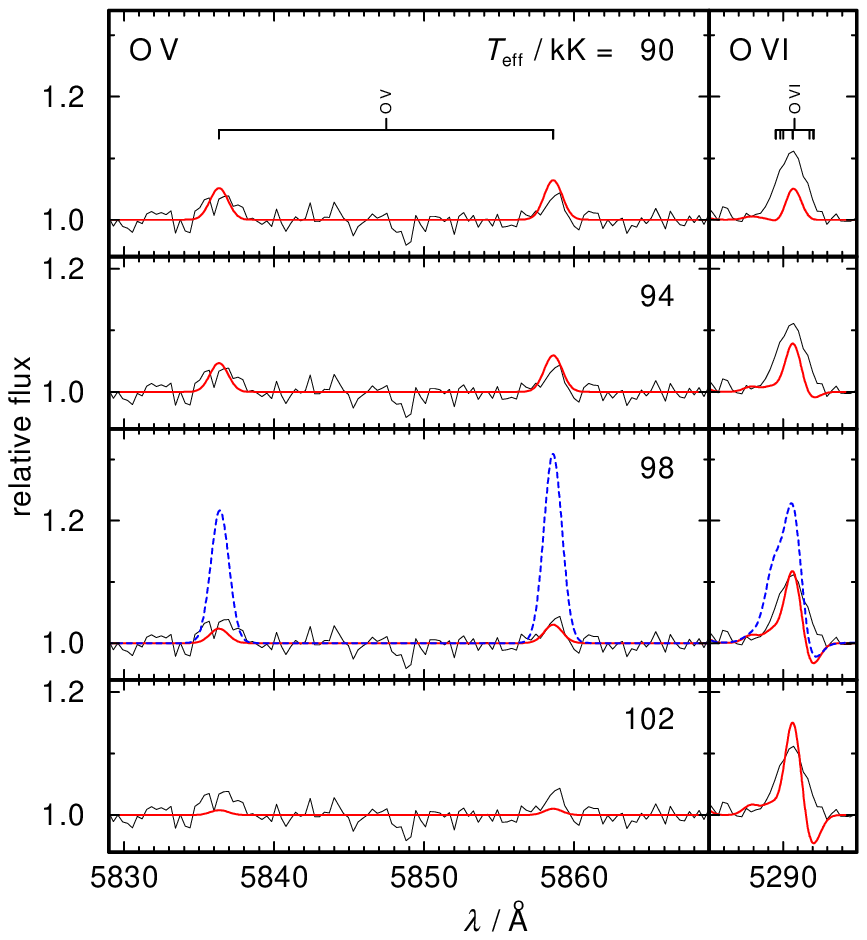}}
  \caption{Theoretical line profiles of \Jonww{O}{v}{5836, 5859} (left) and \Jonw{O}{vi}{5290}
           (right) calculated from H+He+C+N+O models with 
           \Teffw{90\,000, 94\,000, 98\,000, 102\,000} (from top to bottom), \loggw{5.0}
           and an O abundance of $8.0\times 10^{-3}$ (mass fraction; red, full lines).
           The dashed, blue spectrum in the 98\,000\,K panel was calculated from a model with 
           an O abundance of $5.7\times 10^{-3}$ \citep[solar value,][]{asplundetal2009}.
          } 
  \label{fig:TheoSSA_Teff}
\end{figure}

We adjusted \Teff to reproduce the observed equivalent-width ratio of \Jonww{O}{v}{5838, 5860} 
and \Jonw{O}{vi}{5290}. It is best matched at \Teffw{98\,000}. We verified that \loggw{5.0} is
well in agreement with the observation at this \Teff.
To fit the observed \ion{O}{v} und \ion{O}{vi} equivalent widths, 
we had to increase the O abundance by a factor of 1.47 (Fig.\,\ref{fig:TheoSSA_Teff}).

\begin{figure}
  \resizebox{\hsize}{!}{\includegraphics{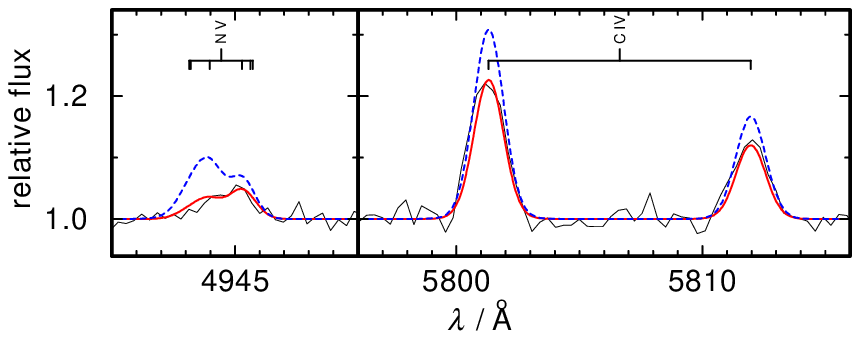}}
  \caption{Left: Theoretical line profiles of 
           \Jonww{N}{v}{4943 - 4944} calculated from H+He+C+N+O models with 
           \Teffw{98\,000}, \loggw{5.0}, and N mass fractions of
            $3.1\times 10^{-4}$ (red, full line) and
            $6.9\times 10^{-4}$ \citep[solar value, blue, dashed line,][]{asplundetal2009}
            compared with the observation (black line).
           Right: Like left panel, for 
           \Jonww{C}{iv}{5801.33, 5811.98}.
           The models were calculated with C mass fractions of 
           $2.0\times 10^{-3}$ (red, full line) and
           $2.4\times 10^{-3}$ \citep[solar value,][]{asplundetal2009}.
          } 
  \label{fig:cn}
\end{figure}

\begin{figure*}
  \resizebox{\hsize}{!}{\includegraphics{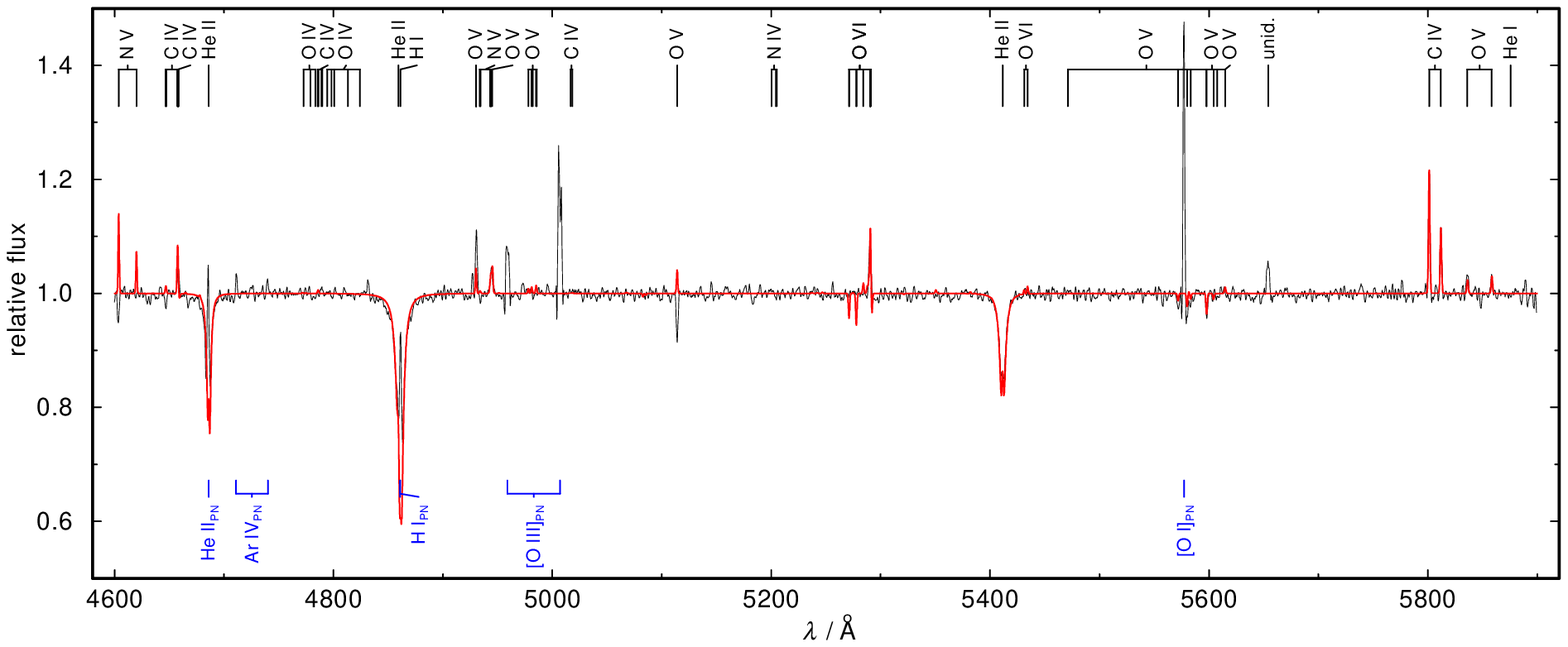}}
  \caption{Same like Fig.\,\ref{fig:TheoSSA_solar}, with adjusted photospheric
          parameters (Table\,\ref{tab:results}).
          } 
  \label{fig:TheoSSA_best}
\end{figure*}

Although the \Teff determination from the \ion{O}{v} / \ion{O}{vi} equilibrium is extremely
sensitive, we estimate a realistic error of 5\,000\,K, that accounts for the uncertainty
of \loggw{5.0^{+0.3}_{-0.2}} and of the photospheric abundances. Opacities that are neglected
in our model calculation may have an additional impact. We verified that \loggw{5.0} is
well in agreement with the observation at \Teffw{98\,000}.
In a last step, we adjusted the
C (decreased by a factor of 0.84) and N 
(decreased by a factor of 0.46)
abundances to improve the agreement between model and the observation.
Example lines of C and N, calculated with solar and our adjusted abundances, are shown in Fig.\,\ref{fig:cn}.

A new model was calculated with the adjusted C, N, and O abundances.
Figure\,\ref{fig:TheoSSA_best} shows its synthetic spectrum.
The previous determinations of \Teff, \logg, and the abundances were verified.
We estimate an error of $\pm 0.2\,\mathrm{dex}$ for the abundances considering the propagation
of the uncertainties of \Teff, \logg, and the background opacities.

We encounter the problem that our best model does not reproduce some lines,
e.g., \Jonw{O}{v}{5114.06}. This line is in agreement with the observed absorption 
line at \Teffw{80\,000} (Fig.\,\ref{fig:TheoSSA_solar})
but appears slightly in emission at \Teffw{98\,000} (Fig.\,\ref{fig:TheoSSA_best}).
This may be a result of the incompleteness and/or the accuracy of the atomic
data in the O model atom that was used to calculate the stellar atmospheres.
Moreover, additional opacities of metals that were not considered in our
models may have an impact on the atmospheric structure. 
Since we used the average observed FORS2 spectrum (Sect.\,\ref{sect:intro})
the spectroscopic variability is another source of uncertainty.
A similar problem was recently reported by \citet{werneretal2014}
in case of the O(He)-type star SDSS\,J172854.34+361958.6 that
has about the same \Teffw{100\,000\pm 10\,000} and \loggw{5.0 \pm 0.2}.
Thus, the investigation of the ``\Jonw{O}{v}{5114.06} problem'' may be subject
of further analyses. However, like already mentioned, this is not the focus of
this paper.

\section{Results and conclusion}
\label{sect:results}

In the framework of the VO, GAVO provides the registered VO 
service TheoSSA and the VO tool TMAW.
These allow to access synthetic stellar SEDs at three levels.

\begin{enumerate}
\item {\sc Fast  and easy.} Download of precalculated SEDs via TheoSSA 
       -- even an unexperienced VO user 
       (no detailed knowledge of the model-atmosphere code necessary) can compare observed
       and synthetic spectra for stellar classification and preliminary determination of
       \Teff, \logg, and photospheric abundances within uncertainties of about 20\,\%.
\item {\sc Individual.} Calculation of SEDs (elements H, He, C, N, O, Ne, Na, and Mg) with standard 
       TMAD model atoms for a preliminary analysis of individual objects with TMAW.
       Improved determination of \Teff, \logg, and abundances within 20\,\%.
\item {\sc Experienced.} Advanced calculation of individual SEDs with own model atoms, e.g\@.
       for the detailed comparison of SEDs calculated with other stellar model-atmosphere codes.
\end{enumerate}

As an example to show the ease of TheoSSA and TMAW usage,
we performed a spectral analysis of the exciting star of \pn,
we successfully employed the GAVO service TheoSSA and the GAVO tool TMAW 
to calculate NLTE stellar atmosphere models and SEDs. These were compared to medium-resolution optical 
observations. In a preliminary, coarse analysis with available SEDs from H+He models (solar abundances)
that were retrieved 
via TheoSSA, we determined \Teffw{80\,000 \pm 15\,000} and \loggw{5.0^{+0.3}_{-0.2}}. 
In a subsequent, more detailed analysis that considered C, N, and O opacities in addition, 
the \Teff determination was improved
by the evaluation of the \ion{O}{v} / \ion{O}{vi} ionization equilibrium.
We arrive at \Teffw{98\,000 \pm 5\,000}.
Our results are summarized in Table\,\ref{tab:results}. 
All SEDs that were calculated for this analysis are available via TheoSSA.

Our analysis with SEDs from H+He models yields almost the same 
$T_\mathrm{eff} \approx 80\,000\,\mathrm{K}$ and $\log g \approx 5.2$
mentioned by \citet{penaetal1990}. This shows that H+He composed models 
(presumably used by M\'endez \& Ruiz as well, cf., Sect\,\ref{sect:intro}),
that neglect metal opacities, are not sufficient for this analysis.

\begin{table*}\centering
\caption{Parameters of the CS of \pn as determined by our analysis.
         Our estimated abundance uncertainty is $\pm 0.2$\,dex.
         Column~6 shows the abundances of the PN for comparison.
         Column~7 to 9 give the final (WD) surface composition (mass fractions) of evolutionary tracks with different initial metalicity $Z$
         and final core mass $M$, respectively
        }         
\label{tab:results}
\setlength{\tabcolsep}{.4em}
\begin{tabular}{rlr@{.}lr@{.}lr@{.}lr@{.}lcr@{.}lcr@{.}lr@{.}lr@{.}l}
\hline
\hline
\noalign{\smallskip}                                                                                                     
\multicolumn{2}{l}{$T_\mathrm{eff}\,/\,$K}             & \multicolumn{2}{l}{$98\,000\pm 5\,000$}   & \multicolumn{15}{l}{}    \\
\multicolumn{2}{l}{$\log\ ( g\,/\,\mathrm{cm/s^2} )$} & \multicolumn{2}{l}{$5.0^{+0.3}_{-0.2}$}     & \multicolumn{15}{l}{}    \\
\noalign{\smallskip}                                                                                          
\hline
\noalign{\smallskip}                                                                                          
&                          & \multicolumn{8}{c}{Central Star}                                                                                                                              & & \multicolumn{2}{c}{PN}                                                    & & \multicolumn{6}{c}{stellar post-AGB evolutionary models$^{\mathrm d}$}                                                     \\
\cline{3-10}                     
\cline{12-13}                     
\noalign{\smallskip}                                                                                          
&                          & \multicolumn{2}{c}{Mass}   & \multicolumn{2}{c}{Number} &  \multicolumn{2}{c}{}                       & \multicolumn{5}{c}{}                                                                                                                & & \multicolumn{2}{c}{$M = 0.657\,M_\odot$} & \multicolumn{2}{c}{$M = 0.833\,M_\odot$} & \multicolumn{2}{c}{$M = 0.710\,M_\odot$}  \\
\cline{3-6}                     
\multicolumn{10}{c}{}                                                                                                                                                    \vspace{-5.5mm}                                                                                                                                                                                                                   \\
& Element                  & \multicolumn{2}{c}{}       & \multicolumn{2}{c}{}       & \multicolumn{2}{c}{~~~[X]$^{\mathrm a}$} & \multicolumn{2}{c}{~$\log \epsilon$$^{\mathrm b}$} & & \multicolumn{2}{c}{~~~~~$\log \epsilon$$^{\mathrm b,c}$} \vspace{-2mm}                                                                                                                                   \\
&                          & \multicolumn{4}{c}{fraction}                            & \multicolumn{2}{c}{}                        & \multicolumn{5}{c}{}                                                                                                                & & \multicolumn{2}{c}{$Z = 0.020$} & \multicolumn{2}{c}{$Z = 0.020$} & \multicolumn{2}{c}{$Z = 0.001$}                           \\
\cline{2-13}                     
\cline{15-20}                     
\noalign{\smallskip}                                                                                   
\smspr & \mmspr H                        & $ 7$&$50\times 10^{-1}$ & $ 9$&$25\times 10^{-1}$ & $  0$&$007$ & $12$&$02$ & & \multicolumn{2}{c}{} & & $ $6&$45\times 10^{-1}$ & $ $6&$27\times 10^{-1}$ & $ 6$&$42\times 10^{-1}$ \\
       & \mmspr He                       & $ 2$&$39\times 10^{-1}$ & $ 7$&$43\times 10^{-2}$ & $ -0$&$018$ & $10$&$93$ & & \multicolumn{2}{c}{} & & $ 3$&$21\times 10^{-1}$ & $ 3$&$33\times 10^{-1}$ & $ 2$&$97\times 10^{-1}$ \\
       & \mmspr C                        & $ 1$&$99\times 10^{-3}$ & $ 2$&$06\times 10^{-4}$ & $ -0$&$074$ & $ 8$&$37$ & & $ <7$&$6 \pm 0.3$    & & $ 1$&$33\times 10^{-2}$ & $ 6$&$22\times 10^{-3}$ & $ 4$&$63\times 10^{-2}$ \\
       & \mmspr N                        & $ 3$&$19\times 10^{-4}$ & $ 2$&$83\times 10^{-5}$ & $ -0$&$337$ & $ 7$&$51$ & & $ <8$&$0 \pm 0.3$    & & $ 3$&$22\times 10^{-3}$ & $ 1$&$59\times 10^{-3}$ & $ 5$&$32\times 10^{-4}$ \\
       & \mmspr O                        & $ 8$&$48\times 10^{-3}$ & $ 6$&$59\times 10^{-4}$ & $  0$&$170$ & $ 8$&$87$ & & $  8$&$4 \pm 0.1$    & & $ 1$&$07\times 10^{-2}$ & $ 1$&$02\times 10^{-2}$ & $ 8$&$82\times 10^{-3}$ \\
\hline

\noalign{\smallskip}                                                                                          
\multicolumn{2}{r}{$d^{\mathrm e}\,/\,\mathrm{kpc}$}    & \multicolumn{2}{l}{$9.3^{+3.0}_{-2.8}$}    & \multicolumn{16}{l}{}    \\
\noalign{\smallskip}                                                                                       
\multicolumn{2}{r}{$z^{\mathrm f}\,/\,\mathrm{kpc}$}    & \multicolumn{2}{l}{$5.6^{+1.8}_{-1.7}$}    & \multicolumn{16}{l}{}    \\
\noalign{\smallskip}                                                                                       
\multicolumn{2}{r}{$M\,/\,M_\odot$}                           & \multicolumn{2}{l}{$0.73^{+0.16}_{-0.15}$} & \multicolumn{16}{l}{}     \\
\noalign{\smallskip}                                                                                            
\multicolumn{2}{r}{$\log\ ( L\,/\,L_\odot )$}                 & \multicolumn{2}{l}{$4.2 \pm 0.4$}         & \multicolumn{16}{l}{}    \\
\noalign{\smallskip}                                                                                       
\multicolumn{2}{r}{$R^{\mathrm g}\,/\,\mathrm{~pc} $}    & \multicolumn{2}{l}{$1.35^{+0.44}_{-0.40}$} & \multicolumn{16}{l}{}    \\
\noalign{\smallskip}
\hline         
\hline
\end{tabular}
\newline
Notes.
$^{\mathrm a}$\,{[X] = log (abundance / solar abundance)},
$^{\mathrm b}$\,{$\sum \epsilon_\mathrm{i} = 12.15$},
$^{\mathrm c}$\,{\citet{penaetal1990}},
$^{\mathrm d}$\,{mass fractions, \citet{millerbertolami2016}},
$^{\mathrm e}$\,{distance calculated following \citet{heberetal1984}, \url{http://astro.uni-tuebingen.de/~rauch/SpectroscopicDistanceDetermination.gif}},
$^{\mathrm f}$\,{height below the Galactic plane},
$^{\mathrm g}$\,{linear radius of the PN}
\end{table*}

\begin{figure}
  \resizebox{\hsize}{!}{\includegraphics{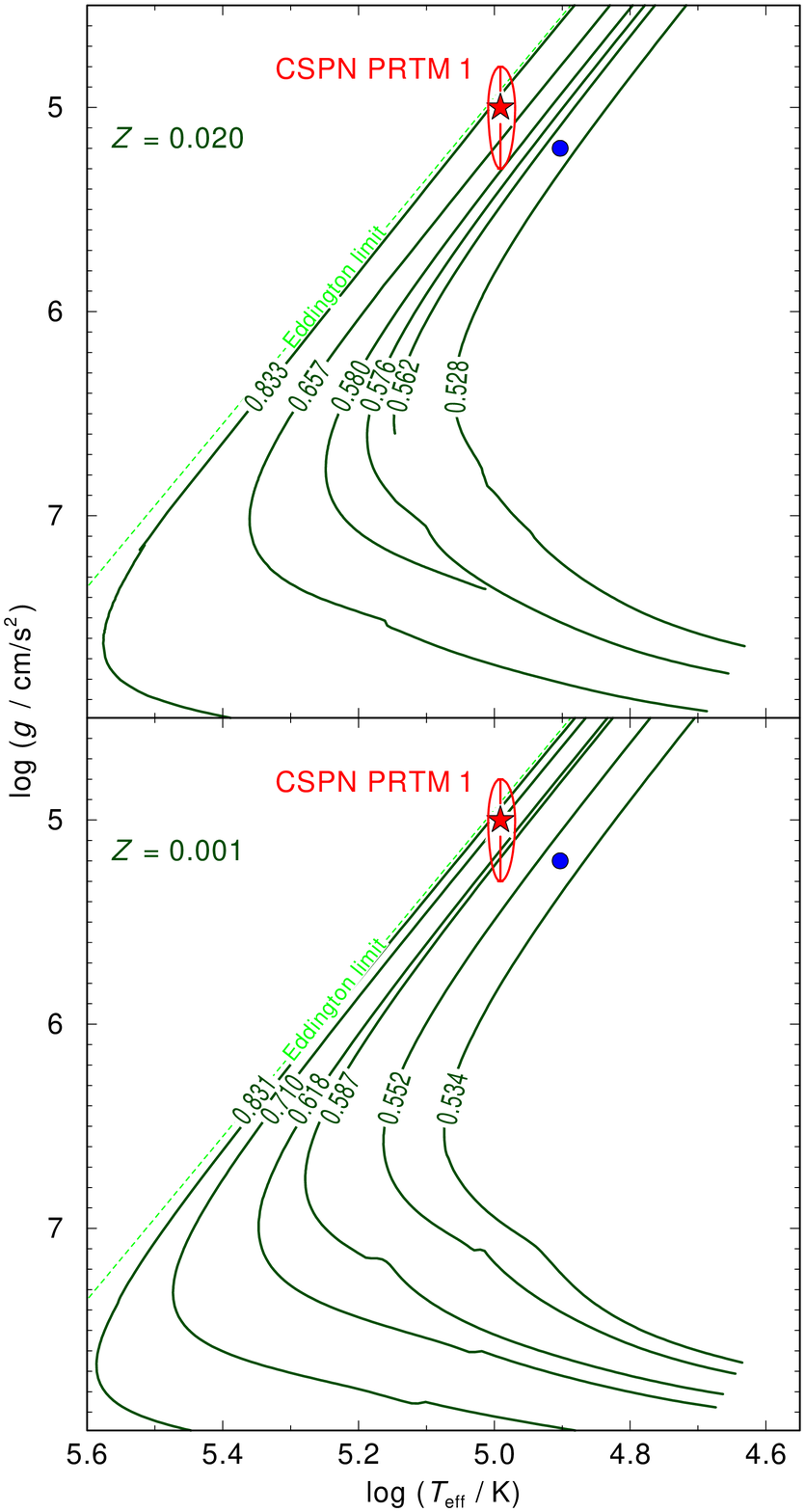}}
  \caption{Location of the CS of \pn in the $\log$ \Teff\,--\,\logg\ diagram.
           Red star: our analysis, the error ellipse is indicated.
           Blue circle: M\'endez \& Ruiz (cf\@. Sect.\,\ref{sect:intro}).
           Post-AGB evolutionary tracks of H-rich stars \citep[for about solar metallicity, $Z = 0.02$,][]{millerbertolami2016}
           labeled with the stellar mass in $M_\odot$, respectively, are shown for comparison. 
           The dashed line indicates the Eddington limit for solar abundances.
          } 
  \label{fig:evolution}
\end{figure}

Figure\,\ref{fig:evolution} shows that the CS of the \pn is located very close to the Eddington limit.
Thus, its cyclic variability may be explained by stellar wind variations \citep[suggested by][]{penaruiz1998} 
or sudden episodic mass-loss events on a slower timescale \citep{feibelman1998}
compared to those discovered in case of the CSPN\,Lo\,4 \citep[PN\,G274.3+09.1,][]{werneretal1992,bond2014}.
These may also be the reason why \citet{feibelman1998} identified 
\ion{O}{vi} emission lines in the ultraviolet (UV) spectra obtained with the International 
Ultraviolet Explorer (IUE) and concluded that the star likely belongs to the so-called
\ion{O}{vi} sequence. \citet{penaruiz1998} did not find \ion{O}{vi} emission lines in the
optical spectra while \citet{boffinetal2012a} clearly found \Jonw{O}{vi}{5290} 
(Figs.\,\ref{fig:TheoSSA_Teff}, \ref{fig:TheoSSA_best}).

For the state-of-the-art spectral analysis of hot stars \citep[cf., e.g.,][]{rauchetal2013,rauchetal2014}, 
high-resolution spectroscopy in the UV wavelength range is necessary, because most of the stellar metal 
lines of highly ionized atoms are located there. These allow to measure ionization equilibria of many species. 
Accurate \Teff and \logg values are a prerequisite for reliable abundance determinations within narrow error
limits. 
In case of the CS of \pn, we neglected its variability and exploited a time-averaged
optical spectrum. We determined important photospheric parameters that allow to investigate on the nature 
of the star. However, time-dependent high-resolution UV spectroscopy would significantly improve the
spectral analysis and is, thus, highly desirable.
 
To determine the stellar mass, we compared \Teff and \logg to stellar evolutionary calculations for H-rich
post-AGB stars that have recently been presented by \citet{millerbertolami2016} (Fig.\,\ref{fig:evolution}). Two
grids with initial metallicities of $Z = 0.020$ (about solar) and $Z = 0.001$ (halo) yield the same mass of 
$M = 0.73 ^{+0.16}_{-0.15}\,M_\odot$. Thus, it is the one of the heavy-weight H-rich post-AGB stars, located close 
to the Eddington limit with a high luminosity of $\log (L\,/\,L_\odot) = 4.2 \pm 0.4$.
The core abundances of the $Z = 0.001$ model agree within error limits with the results of this
spectral analysis (Table\,\ref{tab:results}).
Other models for stellar post-AGB evolution from
\citet{vassiliadiswood1994} and
\citet{bloecker1995}
give similar values for the stellar mass and luminosity like those of \citet{millerbertolami2016}.

We calculated the spectroscopic distance following \citet{heberetal1984} using
$d/\mathrm{pc}=7.11 \times 10^4 \sqrt{H_\nu\cdot M \times 10^{0.4\, m_{\mathrm{v}_0}-\log g}}$
with $m_\mathrm{V_o} = m_\mathrm{V} - 2.175 c$, \citep[$m_\mathrm{V} = 15.6$, $c(\mathrm{H\beta})=0.06$,][]{penaetal1990} and 
the Eddington flux $H_\nu = 1.5 \times 10^{-3}\, \mathrm{erg\, cm^{-2}\, s^{-1}\, Hz^{-1}}$ 
at $\lambda_\mathrm{eff} = 5454\,\mathrm{\AA}$ of our best model atmosphere.
We derived a distance of 
$d=9.3^{+3.0}_{-2.8}$\,kpc 
and a height below the 
Galactic plane of 
$z=5.6^{+1.8}_{-1.7}$\,kpc. 
This agrees within error limits with the previously determined values of
\citet[$d = 5 \pm 2\,\mathrm{kpc}$, $|z| = 3 \pm 1.2\,\mathrm{kpc}$]{penaetal1990}
and verifies that \pn is a halo object.

From the apparent PN diameter of $60\arcsec$ \citep{boffinetal2012a}, we derive its linear radius of 
$R = 1.35^{+0.44}_{-0.40}\,\mathrm{pc}$. \citet{boffinetal2012a} measured an expansion velocity of 
$v_\mathrm{exp} = 30 \pm 5\,\mathrm{km/s}$, that yields an expansion time of $44\,000^{+25\,000}_{-17\,500}\,\mathrm{a}$. 
This is much longer than the post-AGB time of about $120^{+3000}_{\,\,-100}\,\mathrm{a}$, interpolated from evolutionary 
calculations of \citet[$Z = 0.02$ and $Z = 0.001$,][]{millerbertolami2016}. 

The smaller distance determined by \citet{penaetal1990} would slightly reduce the discrepancy by a factor of 0.5, however,
following \citet{millerbertolami2016}, a 0.73\,\Msol\ post-AGB star evolves about two orders of magnitude faster
after its descend from the AGB.
In addition to the huge discrepancy in the theoretical and observed dynamical time scales,
Table\,\ref{tab:results} shows that the
final surface abundances of these models do not agree with the result of our spectral analysis. This and the aged, 
C-poor PN surrounding its relatively massive CS corroborate a presumption that the evolution of the CS of the \pn
cannot be explained with a canonical H-rich single star scenario. A late thermal pulse (LTP) that occurred after the 
first descent from the AGB at still high luminosity, might be responsible for the about solar C and O abundances 
(Table\,\ref{tab:results}), i.e., enriched C and O in this Halo star. 

However, presently no alternative stellar evolutionary models are available that can consistently explain the star's 
location in the $\log$ \Teff\,--\,\logg\ plane, its surface composition, and its post-AGB age.
To summarize, the evolution of the CS is not understood and needs further investigation 
but this is not the focus of this paper.

Based on our higher \Teff and $L$ values, a PN reanalysis with an improved ionizing spectrum appears promising 
to measure the PN abundances (Table\,\ref{tab:results}) again. 

The time to calculate synthetic spectra via TMAW to achieve these results was of the order of three weeks. 
Therefore, all authors and referees should be aware that nowadays possibilities exist for an at least
preliminary spectral analysis by means of NLTE model-atmosphere techniques.
Any analysis, especially all subsequent interpretations and conclusions, would strongly benefit 
if such VO services and tools were used.

\section*{Acknowledgements}
We thank Henri Boffin and Brent Miszalski who put the mean FORS2 spectrum of the CS\pn at our disposal. 
The GAVO project 
had been supported by the Federal Ministry of Education and Research (BMBF) 
at T\"ubingen (05\,AC\,6\,VTB,                 05\,AC\,11\,VTB) and 
at Heidelberg (05\,AC\,6\,VHA, 05\,A0\,8\,VHA, 05\,AC\,11\,VHA, 05\,AC\,14\,VHA).
We thank the GAVO and AstroGrid-D (\url{http://www.gac-grid.org}) teams for support,
especially Ulrike Stampa who created the initial version of TheoSSA.
The TheoSSA service (\url{http://dc.g-vo.org/theossa}) used to retrieve theoretical spectra and
the TMAD service (\url{http://astro-uni-tuebingen.de/~TMAD}) used to compile atomic data for this paper 
were constructed as part of the activities of GAVO.
Based on data products from observations made with ESO Telescopes at the La Silla Paranal Observatory
under programme IDs 078.D$-$0033(A) and 088.D$-$0750(B). 
This research has made use 
of NASA's Astrophysics Data System and
of the SIMBAD database and
   the VizieR catalogue access tool 
\citep[the original description of the VizieR service was published in][]{ochsenbeinetal2000}, 
operated at CDS, Strasbourg, France.




\bibliographystyle{mnras}
\bibliography{theossa} 



\appendix

\section{TheoSSA TMAP Web Interface and TMAW WWW Request Web Interface}

\begin{landscape}

\begin{figure*}
   \includegraphics[trim = 0mm 0mm 0mm 0mm, clip, width=24cm, angle=0]{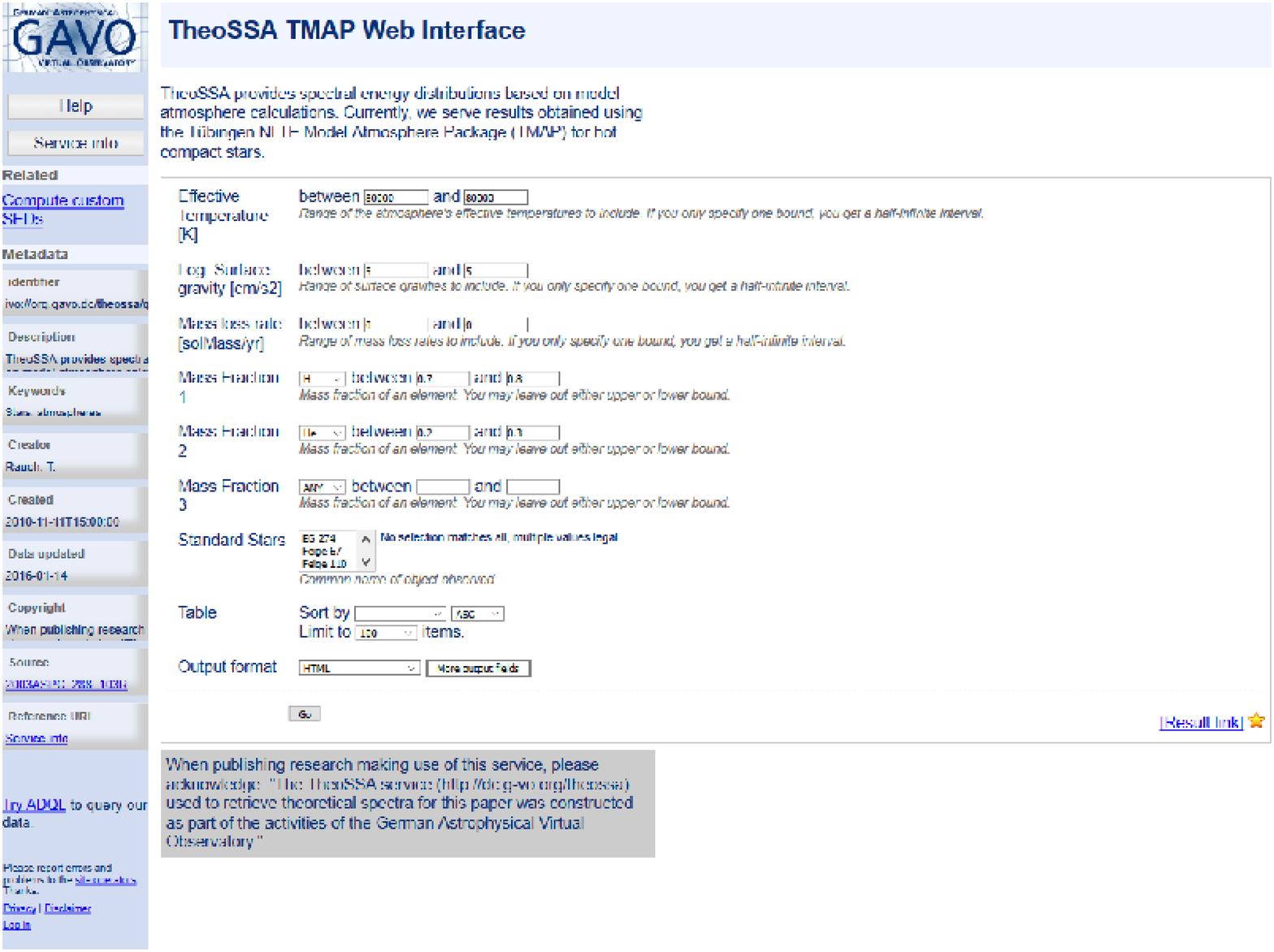}
  \caption{TheoSSA WWW interface.} 
  \label{fig:TheoSSA_input}
\end{figure*}

\end{landscape}

\begin{landscape}

\begin{figure*}
  \includegraphics[trim = 0mm 0mm 0mm 0mm, clip, width=24cm, angle=0]{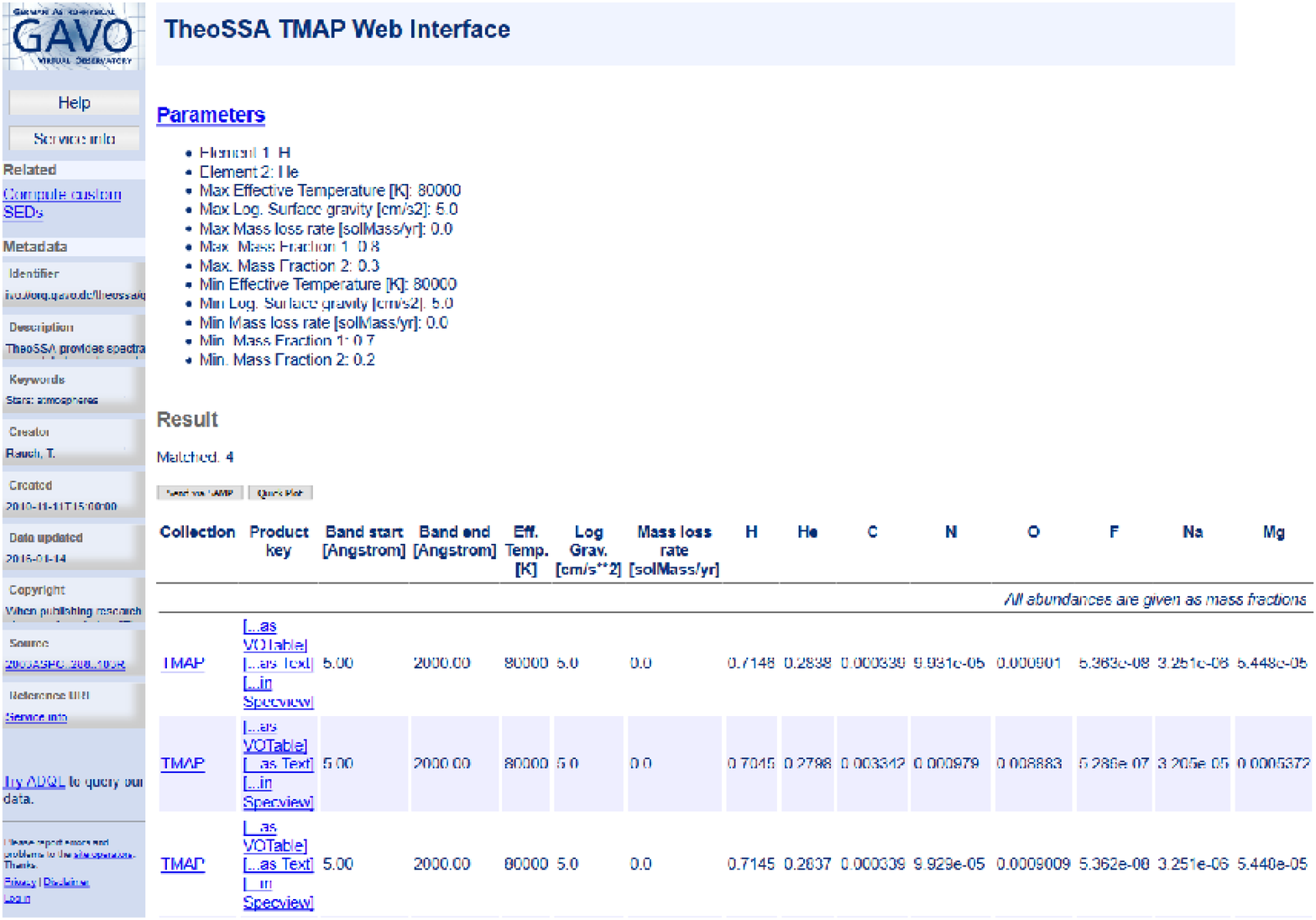}
  \caption{Results of the TheoSSA request shown in Fig.\,\ref{fig:TheoSSA_input}.} 
  \label{fig:TheoSSA_output}
\end{figure*}

\end{landscape}

\begin{landscape}

\begin{figure*}
  \includegraphics[trim = 0mm 0mm 0mm 0mm, clip, width=24cm, angle=0]{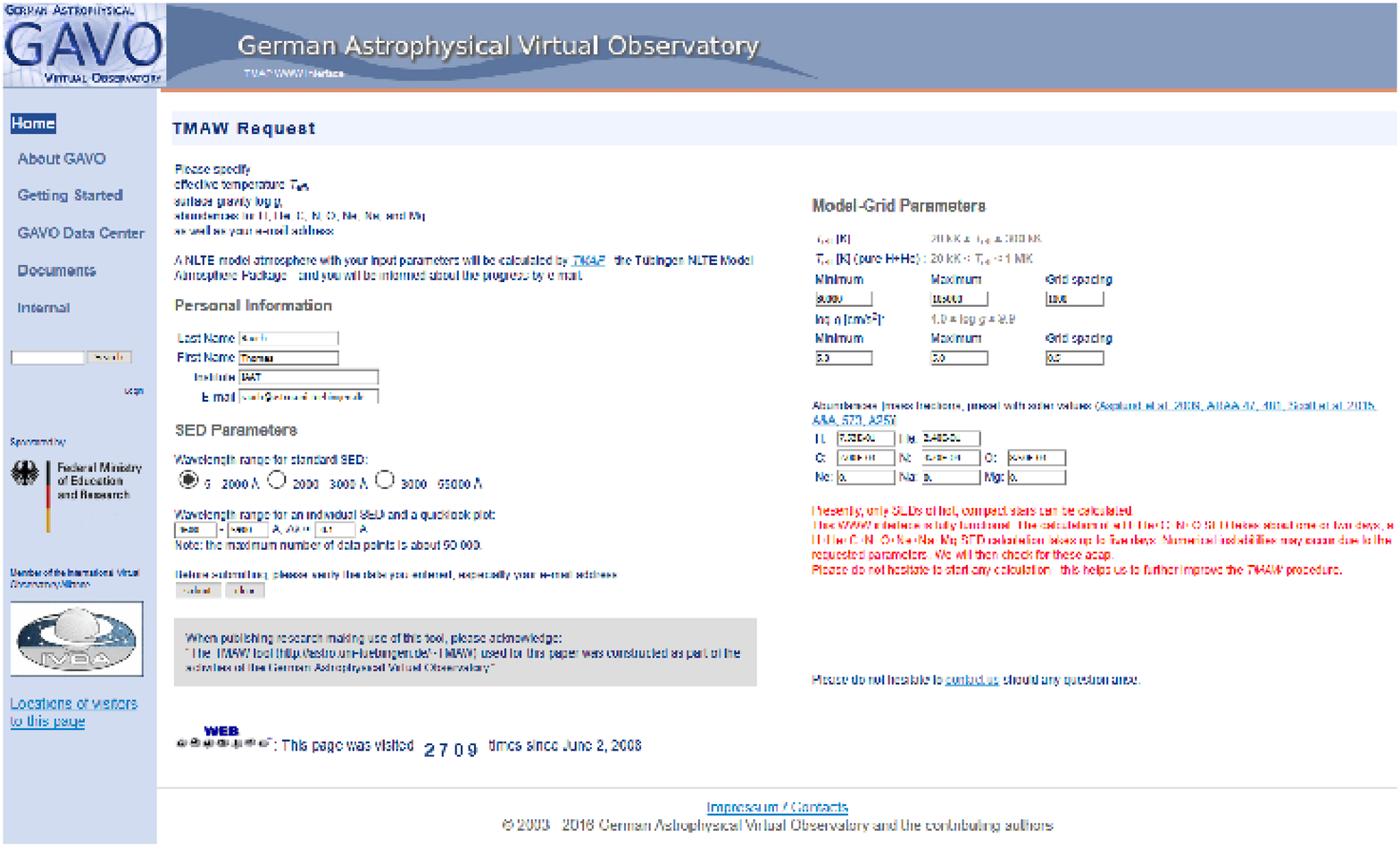}
  \caption{TMAW WWW interface.} 
  \label{fig:TMAW_input}
\end{figure*}

\end{landscape}

\begin{landscape}

\begin{figure*}
  \includegraphics[trim = 0mm 0mm 0mm 0mm, clip, width=24cm, angle=0]{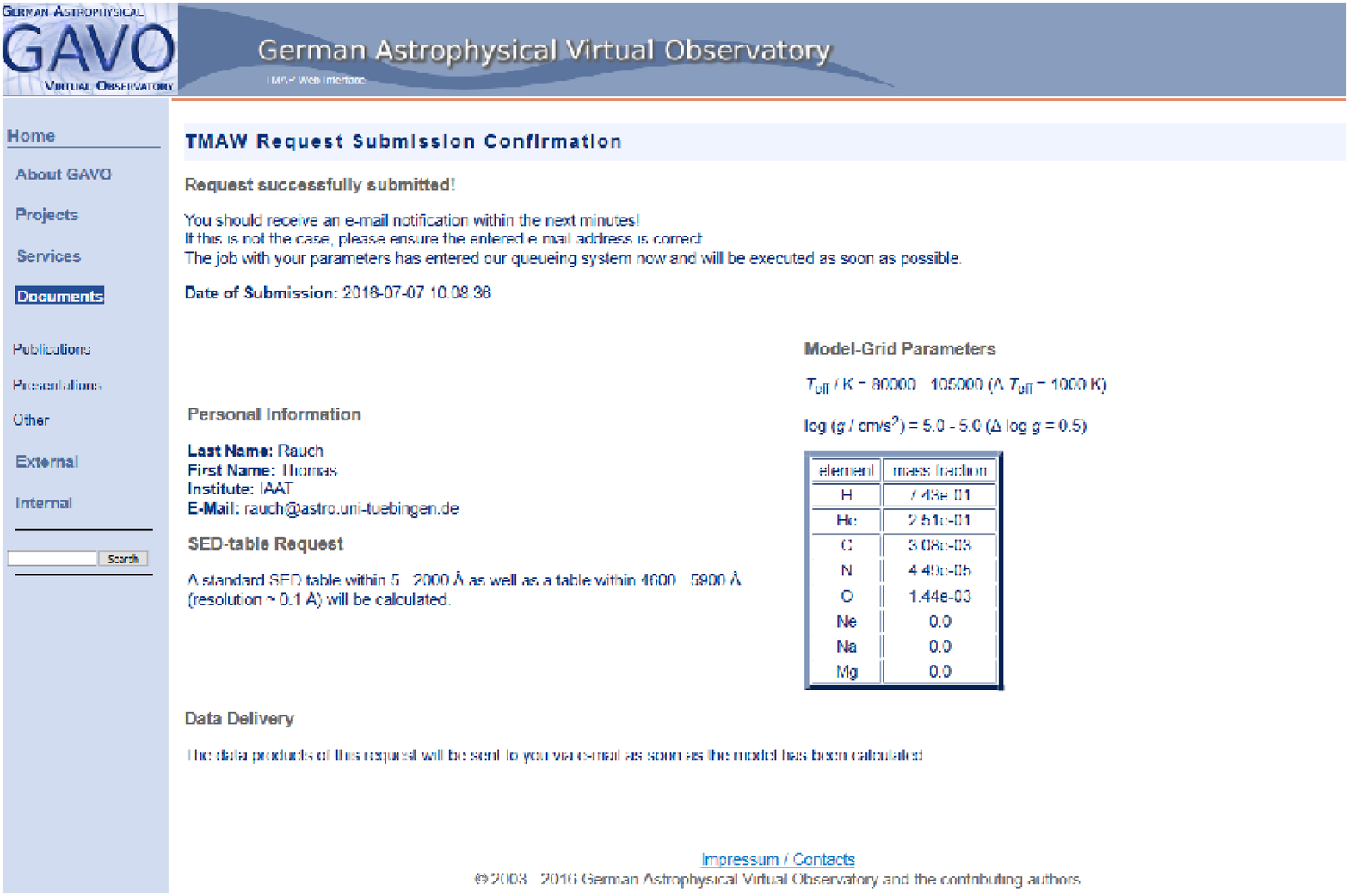}
  \caption{Notification of the successfully submitted TMAW request of Fig.\,\ref{fig:TMAW_input}.} 
  \label{fig:TMAW_output}
\end{figure*}

\end{landscape}

In general, astronomers are busy and do not have time enough to read extensive manuals
or to go through tutorials. Figures \ref{fig:TheoSSA_input} to \ref{fig:TMAW_output} 
demonstrate how the concept of the VO, to be ``easy and intuitive'', has consequently
been implemented in the cases of TheoSSA and TMAW. All input fields in Figs\@. \ref{fig:TheoSSA_input} 
and \ref{fig:TMAW_input} are clearly and unambiguously labeled and, thus, can easily be filled in.

Table\,\ref{lst:request} shows in plain text an example request file that contains all information 
used to calculate the NLTE model atmosphere and subsequently the synthetic spectrum.

Table\,\ref{lst:results} displays the result files that can be retrieved by a TMAW user.
The table files' headers describe their content clearly.

\clearpage
\onecolumn

\clearpage

\clearpage

\clearpage

\newpage

\begin{table}
\caption{Example of a TMAW request file.}
\label{lst:request}
\begin{lstlisting}
5.000
98000
rauch@astro.uni-tuebingen.de                                                    
Thomas                                                                          
Rauch                                                                           
2016-07-22_14_23_36                                                             
7.52E-01                                                                        
2.40E-01                                                                        
2.00E-03                                                                        
3.20E-04                                                                        
8.50E-03                                                                        
0.                                                                              
0.                                                                              
0.                                                                              
4600                                                                            
5900                                                                            
0.1                                                                             
UV                                                                              
\end{lstlisting}
\end{table}

\begin{table}
\caption{Example of TMAW product files.}
\label{lst:results}
{\footnotesize
\begin{tabular}{l}
0098000\_5.00\_\_\\~~~H\_\_7.499E-01\_HE\_2.393E-01\_C\_\_1.994E-03\_N\_\_3.191E-04\_O\_\_8.476E-03\_NE\_0.000E+00\_NA\_0.000E+00\_MG\_0.000E+00\\~~~\_LF\_2016-07-22\_14\_23\_36.flux$^{\mathrm a}$ \vspace{2mm}\\
0098000\_5.00\_\_\\~~~H\_\_7.499E-01\_HE\_2.393E-01\_C\_\_1.994E-03\_N\_\_3.191E-04\_O\_\_8.476E-03\_NE\_0.000E+00\_NA\_0.000E+00\_MG\_0.000E+00\\~~~\_LF\_2016-07-22\_14\_23\_36\_flux.ps$^{\mathrm b}$ \vspace{2mm}\\
0098000\_5.00\_\_\\~~~H\_\_7.499E-01\_HE\_2.393E-01\_C\_\_1.994E-03\_N\_\_3.191E-04\_O\_\_8.476E-03\_NE\_0.000E+00\_NA\_0.000E+00\_MG\_0.000E+00\\~~~\_LF\_2016-07-22\_14\_23\_36.IDENT$^{\mathrm c}$ \vspace{2mm}\\
0098000\_5.00\_\_\\~~~H\_\_7.499E-01\_HE\_2.393E-01\_C\_\_1.994E-03\_N\_\_3.191E-04\_O\_\_8.476E-03\_NE\_0.000E+00\_NA\_0.000E+00\_MG\_0.000E+00\\~~~\_LF\_2016-07-22\_14\_23\_36.TMAW\_out$^{\mathrm d}$ \vspace{2mm}\\
0098000\_5.00\_\_\\~~~H\_\_7.499E-01\_HE\_2.393E-01\_C\_\_1.994E-03\_N\_\_3.191E-04\_O\_\_8.476E-03\_NE\_0.000E+00\_NA\_0.000E+00\_MG\_0.000E+00\\~~~\_LF\_2016-07-22\_14\_23\_36\_T-structure.ps$^{\mathrm e}$ \vspace{2mm}\\
0098000\_5.00\_\_\\~~~H\_\_7.499E-01\_HE\_2.393E-01\_C\_\_1.994E-03\_N\_\_3.191E-04\_O\_\_8.476E-03\_NE\_0.000E+00\_NA\_0.000E+00\_MG\_0.000E+00\\~~~\_LF\_2016-07-22\_14\_23\_36.WFP$^{\mathrm f}$ \vspace{2mm}\\
0098000\_5.00\_\_\\~~~H\_\_7.499E-01\_HE\_2.393E-01\_C\_\_1.994E-03\_N\_\_3.191E-04\_O\_\_8.476E-03\_NE\_0.000E+00\_NA\_0.000E+00\_MG\_0.000E+00\\~~~\_LF\_2016-07-22\_14\_23\_36.WFP\_00005-02000$^{\mathrm g}$ \\
\end{tabular}
}
\newline
Notes.
$^{\mathrm a}$\,{Flux of the model atmosphere,}
$^{\mathrm b}$\,{plot of~ $^{\mathrm a}$,}
$^{\mathrm c}$\,{wavelengths of lines included in the calculation of the synthetic spectrum (user chosen wavelength range and resolution),}
$^{\mathrm d}$\,{output of the model-atmosphere calculation,}
$^{\mathrm e}$\,{plot of the temperature structure of the model atmosphere,}
$^{\mathrm f}$\,{synthetic spectrum (user chosen wavelength range) with 
                  $\lambda$, 
                  astrophysical flux $F_\mathrm{\lambda}$, and 
                  rectified flux $F_\mathrm{\lambda}/F_\mathrm{\lambda}^\mathrm{continuum}$,}
$^{\mathrm g}$\,{same as~ $^{\mathrm f}$ but for a standard wavelength range (here 5 to 2000\,\AA.}
\end{table}
\

\bsp	
\label{lastpage}
\end{document}